\documentclass[manuscript]{aastex}
\usepackage{epsfig}
\newcommand{\be}{\begin{equation}}
\newcommand{\ee}{\end{equation}}
\newcommand{\bea}{\begin{eqnarray}}
\newcommand{\eea}{\end{eqnarray}}
\newcommand{\etal}{\mbox{\rm et al.~}}
\newcommand{\ms}{\mbox{m s$^{-1}~$}}
\newcommand{\kms}{\mbox{km s$^{-1}~$}}

\newcommand{\mse}{\mbox{m s$^{-1}$}}

\newcommand{\msun}{M$_{\odot}~$}
\newcommand{\msune}{M$_{\odot}$}

\newcommand{\lsune}{L$_{\odot}$}
\newcommand{\mjup}{M$_{\rm JUP}~$}
\newcommand{\mjupe}{M$_{\rm JUP}$}

\newcommand{\mearth}{M$_{\rm Earth}~$}
\newcommand{\mearthe}{M$_{\rm Earth}$}

\newcommand{\msini}{$M \sin i~$}

\newcommand{\scs}{$\sqrt{\chi^2_{\nu}}$}
\newcommand{\arel}{$a_{\rm rel}$}

\received{}
\accepted{}
\slugcomment{ Accepted to PASP: 07 Nov 2006}
\shortauthors{Maness {\it et~al.\/}}
\shorttitle{Modeling the Neptune--Mass Host, GJ 436}
\begin{document}
\title{~~\\ ~~\\ The M Dwarf GJ 436 and its
  Neptune-Mass Planet}
\author{ H. L. Maness\altaffilmark{1}, {G. W. Marcy\altaffilmark{1}},
{E. B. Ford\altaffilmark{1}},  
{P. H. Hauschildt\altaffilmark{2}}, {A. T. Shreve\altaffilmark{3}}, 
{G. B. Basri\altaffilmark{1}}, {R. P. Butler\altaffilmark{4}},  
{S. S. Vogt\altaffilmark{5}}}
\altaffiltext{1}{Department of Astronomy, University of California at 
Berkeley, Berkeley, CA, USA 94720}
\altaffiltext{2}{Hamburger Sternwarte, Gojenbergsweg 112, 21029
Hamburg, Germany}
\altaffiltext{3}{College of Chemistry, University of California at
  Berkeley, Berkeley, CA, USA 94720}
\altaffiltext{4}{Department of Terrestrial Magnetism, Carnegie
Institution of Washington, 5241 Broad Branch Rd NW, Washington DC, USA
20015-1305}
\altaffiltext{5}{UCO/Lick Observatory, University of California at
Santa Cruz, Santa Cruz, CA, USA 95064}

\begin{abstract}
We determine stellar parameters for the M dwarf GJ~436 that hosts a
Neptune-mass planet.  We employ primarily spectral modeling at low and high 
resolution, examining the agreement
between model and observed optical spectra of five comparison stars of
type, M0-M3.  Modeling high resolution optical spectra suffers from
uncertainties in TiO transitions, affecting the predicted strengths of
both atomic and molecular lines in M dwarfs.  The determination of
T$_{\rm eff}$, gravity, and metallicity from optical spectra remains at
$\sim$10\%.  As molecules provide opacity both in lines and as an
effective continuum, determing molecular transition parameters remains
a challenge facing models such as the PHOENIX series, best verified
with high resolution and spectrophotometric spectra.  Our analysis of GJ 436
yields an effective temperature of T$_{\rm eff}$ = 3350 $\pm$ 300 K
and a mass of 0.44 \msune.  New Doppler measurements for
GJ~436 with a precision of 3 \ms taken during 6 years improve the 
Keplerian model of the planet, giving a minimum mass, $M\sin i$ =
0.0713 \mjup = 22.6 \mearth, period, $P$ = 2.6439 d, and $e$ =
0.16$\pm$0.02.  
The noncircular orbit contrasts with the tidally
circularized orbits of all close-in exoplanets, implying either
ongoing pumping of eccentricity by a more distant companion, or
a higher Q value for this low-mass planet.  The velocities indeed reveal
a long term trend, indicating a possible distant companion.
\end{abstract}

\pagebreak


\section{Introduction}

To date, radial velocity surveys have revealed three exoplanetary systems 
with M dwarf hosts: GJ 876, GJ 436, and GJ 581 \citep{Rivera05,
Butler04,Bonfils05b}.  Remarkably, all three of these M dwarfs host
planets with minimum masses (\msini) less than 0.1 \mjupe, ranking
them among the lowest mass exoplanets known. Microlensing surveys have 
strengthened the case for the preferntial occurence of such ``super-Earths'' 
around M dwarfs.  Current results suggest four M dwarf systems, with two 
harboring planets in the super-Earth mass range \citep{Beaulieu06,Gould06,
Bond04,Udalski05}.  Given detection biases against the discovery of 
low mass planets, all these findings imply that super-Earths are more 
common close companions to M dwarfs than gas giants \citep{Endl06,Beaulieu06}.
The frequency of longer period planets remains poorly constrained.

From a theoretical viewpoint, the growth of planets around low mass
stars has been explored only recently.  Low mass stars likely form by
accretion at the centers of protoplanetary disks having lower mass
than those surrounding solar mass stars.  Such low mass disks may
spawn low mass planets, both because of less available mass and
because of shorter survival times for the disks (Laughlin et al
2004).  However, core accretion models of giant planet formation that include 
growth of a rocky core from dust particles followed by the gravitational
accretion of gas suggest that giant planets may have enough time to
form. Solutions to the time scale competition between planet growth
and disk lifetime have been proposed by invoking both migration to
move the planet to gas-rich areas and also by adopting lower opacities
to allow faster accretion of the envelope onto the core (Hubickyj et
al. 2004, Alibert et al. 2005).  Nonetheless, rocky cores that are
starved of gas may instead become ice giants similar to Neptune and
Uranus (Goldreich et al. 2004, Thommes et al. 2002, Ida \& Lin 2005).  
Under the core accretion paradigm, it remains unknown how commonly 
planets fail to accrete gas after successful 
growth of a rock and ice core, thereby leaving behind a super-Earth planet.
Most recently, \citet{Boss06} provided a very different viable 
alternative to the core accretion scenario, showing that disk instability 
can also be invoked to explain current M dwarf observations.  Distinguishing
between these competing models of planet formation requires more observational results.

To further understanding in this area, 
we present a follow-up study of GJ 436, an M dwarf recently discovered to host a 
Neptune-mass planet (\msini = 0.066 \mjup; Butler et al. 2004).  At present, little 
is known about this system.  As there is 
no other giant planet within a few AU of the host star, a larger 
planet cannot easily be invoked to explain a curtailed accretion of gas of 
the known planet, hereafter GJ~436b.  Furthermore, since the stellar luminosity and 
mass are not well known, the planet's minimum mass and predicted 
temperature are poorly constrained.  
To better characterize this system, we have obtained new Doppler measurements 
and attempted to deduce accurate stellar parameters.  
In \S 2, we discuss photometric observations leading to estimates
of the stellar mass and luminosity.  In \S 3, we present an effort to
determine the effective temperature, surface gravity, and
metallicity of GJ 436 via low and high resolution spectral modeling.
In \S 4, new Doppler measurements are presented, leading to an
improved orbit and minimum mass for the planet.  In \S 5 and 6, 
we discuss the implications of our results.

\section{Photometric Observations and Analysis}
  
We first assess the stellar properties of GJ~436 using photometric
measurements and published calibrations.  We use optical and
near-infrared photometry for GJ 436 taken from the
compilation by \citet{Leggett92}.  The quoted values are V = 10.66, I
= 8.28, J = 6.93, H = 6.34, and K = 6.10.  We adopt the Hipparcos parallax
for GJ 436 of 97.73 mas \citep{Perryman97}.

\subsection{Stellar Mass}

We estimate the mass of GJ 436 from various empirical mass-luminosity 
relations (MLRs) and theoretical models for M dwarfs.
\citet{Delfosse00} have determined empirical MLRs for visual and 
near-infrared magnitudes using their newly discovered M dwarf 
binaries.  With these relations, the V, J, H, and K band photometry 
from \citet{Leggett92} yield
inferred masses for GJ~436 of 0.418, 0.439, 0.441, and 0.442 \msune,
respectively.  Thus, the JHK photometric calibrations all yield a
stellar mass of M${_\star}$ = 0.44 ${\pm}$ 0.04 M${_\sun}$.  We adopt
an accuracy of $\sim$10\% in stellar mass from the scatter in the
calibrations of \citet{Delfosse00}.  The V band estimate of 0.418
\msun is lower than the mass derived from the the JHK calibration, but
the V-band calibration has more scatter.  Moreover, the particular
metallicity of GJ~436 will affect the mass estimate from the V band
more than from the IR bands.  As metallicity remains somewhat
uncertain (\S 3), we give more weight to the mass from the near-IR
calibrations.

For comparison, the mass estimates derived from both the empirical
relations of \citet{Henry93} and \citet{Benedict01} agree within 10\%
of 0.44 \msune.  The theoretical models of \citet{Baraffe98} and
\citet{Siess00} applied to GJ~436 also yield masses that are in
agreement at the 10\% level, though the theoretical relation between mass
and optical flux has a strong metallicity dependence.  Here, we adopt the 
mass derived from the near-infrared
relations of \citet{Delfosse00}, giving M${_\star}$ = 0.44 ${\pm}$
0.04 M${_\sun}$.  This value is 7.3\% higher than that adopted by
Butler et al. (2004), who gave more weight to the mass derived from the
various V band calibrations.

\subsection{Stellar Luminosity}

Extensive observational analyses of M dwarf luminosities have been
performed by \citet{Leggett96}, who derived luminosities for 16 M
dwarfs by combining spectrophotometry and broadband measurements over
the wavelength range, 0.35 to 5 $\mu$m.  \citet{Reid00} have fit
second-order polynomials to the derived bolometric corrections of
\citet{Leggett96} in the V, I, and K bands.  Employing these
relations for GJ~436 yields 
implied luminosities of 0.024, 0.024, and 0.025 $L_{\sun}$,
respectively.

These estimates are in good agreement with those derived from the
tight empirical relation between M$_K$ and M$_{BOL}$ of
\citet{Veeder74}.  The observed relations for luminosity are also in
agreement with those from theoretical models constructed by
\citet{Baraffe98}, \citet{Siess00}, and \citet{Dantona94}.  Here, 
we assume the simple average of the derived luminosities quoted
above: $L = 0.024 \pm 0.004 L_{\sun}$, slightly lower than the value,
$L$=0.025 \lsune, that was found by Butler et al. (2004).

\section{Spectral Modeling}

\subsection{Model Atmospheres}

To constrain the effective temperature, metallicity, and surface
gravity of GJ 436, we compared observed spectra obtained at both low
and high resolution to synthetic spectra.  We used an updated grid of
the NEXTGEN set of M dwarf models \citep{Hauschildt99} that includes
an updated molecular line list, a revised treatment of dust grain
formation, and a treatment of spherical geometry.  A detailed
description of the model atmospheres used in calculating the synthetic
spectra presented here can be found in \citet{Allard01}.

\subsection{High Resolution Modeling}

We obtained high resolution (R $\sim$ 60,000) echelle spectra for
GJ~436 and five comparison M dwarfs (GJ 411, GJ 424, GJ 752A, GJ 860
A, and GJ 908) using the Keck 1 telescope with the HIRES spectrometer
between June 1997 and November 2001.  The wavelength range was
3700-6200 \AA, contained in 33 spectral orders. Only spectra at
wavelengths greater than 5203 \AA \ were used in spectral modeling.  The
standard planet-hunting procedure for reduction of the raw echelle
images was employed to perform flat-fielding, sky subtraction, order
extraction, and wavelength calibration \citep{Butler96, Vogt94}.  All 
spectral modeling was performed on template spectra not containing 
iodine.

To compare the observations to the models, several adjustments were
made to the continuum and wavelength scale of the observed spectra.
The pseudo-continuum levels of the M dwarf spectra consist of
overlapping molecular bands, and the echelle spectra are not flux
calibrated.  To bring the models and the observed spectra to the same
continuum value, we fit a second degree polynomial to the points of
highest flux across each echelle order and also to the corresponding
wavelength section of the model spectrum.  We then divided by these
pseudo-continuum polynomials to obtain normalized fluxes for both
observed and synthetic spectra.  We next degraded the resolution of
the model spectra to match the observed resolution by convolving each
model spectrum with a Gaussian of the appropriate width.  It was not
necessary to modify the models further to correct for rotational line
broadening, as none of the stars in our sample have significant
rotation, having V$\sin i <$ 3 \kms \citep{Delfosse98}.  Finally, we
removed the stellar Doppler shift from the observed spectra by 
cross-correlating 
the observed and model spectra and shifting the observed
spectra by the appropriate amount.

Comparing observed to synthetic spectra constitutes a standard
approach in determining M dwarf stellar parameters \citep{Jones94,
Leggett96, Valenti98, Basri00, Leggett00, Leggett01, Leggett02,
Kirkpatrick93, Bean06}.  However, deducing characteristics of early type M
dwarfs using high resolution spectral modeling has been performed
rarely and has revealed uncertainties with the molecular constants
\citep{Valenti98}.  To derive characteristics of GJ~436 from high
resolution spectral modeling, we therefore tested the integrity of our
model fits by tests on other early type M dwarfs.  The comparison M
dwarfs were selected to encompass the range of early spectral types
and metallicities near GJ~436.

\begin{figure}
\centerline{\scalebox{0.9}{\plotone{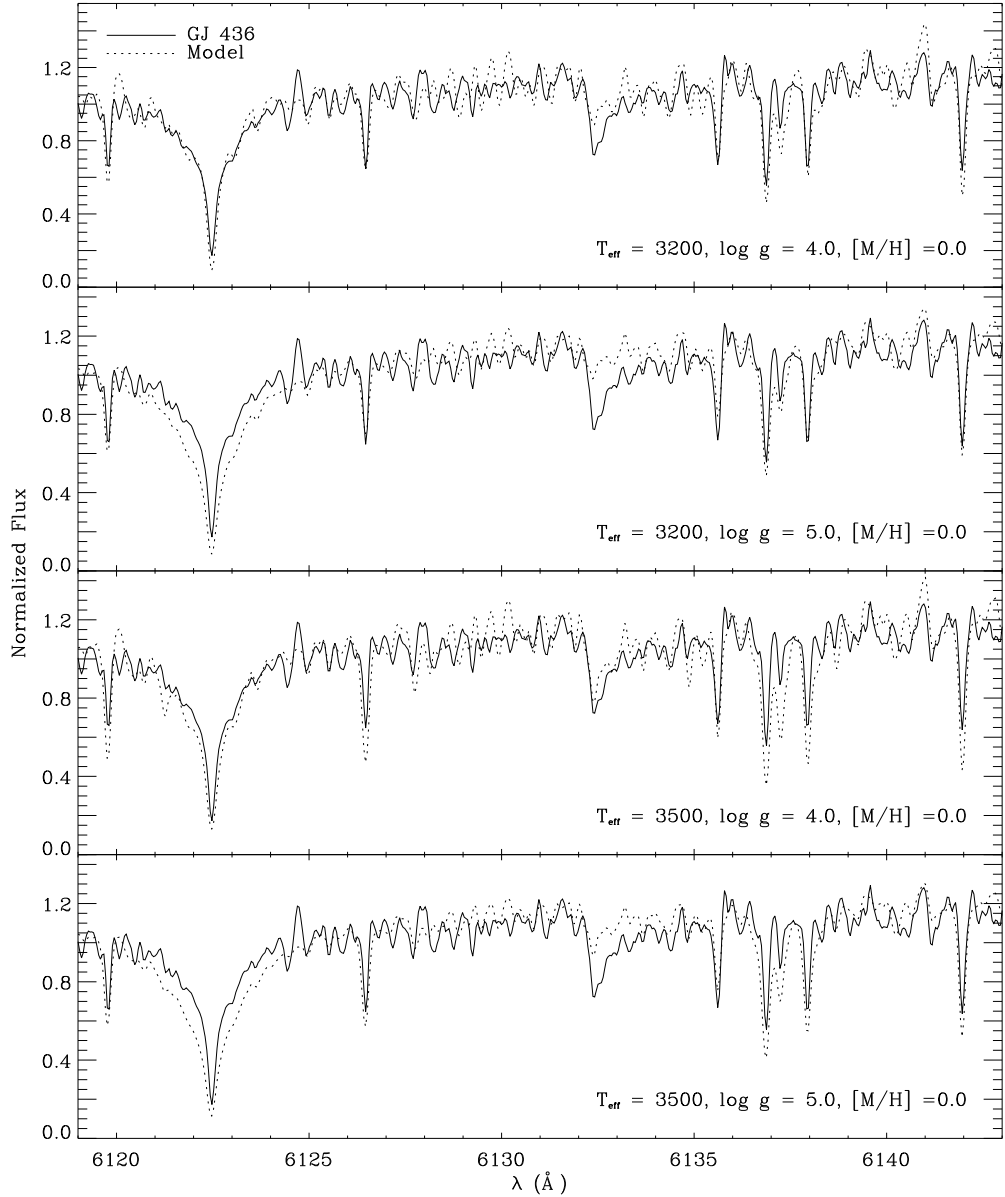}}}
\caption{Comparison at high resolution of the high 
resolution best fit and the low resolution best fit for GJ 436. 
The figures show a representative portion of one echelle order; notice 
that in all figures, the model TiO molecular lines do not match the 
observed lines.  The top panel shows the best fit at this resolution.  
However, while the atomic line profiles are well fit, and the 
overall strength of the TiO lines is similar, the surface
gravity for this model is unphysical (see the text).  When the surface
gravity is fixed to a reasonable value (panels 2 and 4), the overall 
strength of the molecular lines and the atomic line profiles can not be
simultaneously matched.}
\label{fig1}
\vspace{10mm}
\end{figure}

The results of our high resolution modeling of GJ~436 are summarized
in Figure \ref{fig1}, showing overplotted synthetic and observed
spectra for a representative portion of one echelle order.  Figure
\ref{fig1} demonstrates that the model TiO lines, constituting a
pseudo-continuous opacity for M dwarfs in the optical, do not match
the observed molecular lines.  This discrepency was observed in all
orders and for all M dwarfs in our sample.  The specific wavelengths
and pattern of wavelengths of the observed TiO lines is so different
from those in the synthetic spectrum that no association of the two
sets of TiO lines was possible.  The apparent flaws in the molecular
constants for TiO are reminiscent of those described for TiO by
\citet{Valenti98} in their high resolution spectra of M dwarfs.
Apparently, the treatment of TiO at high resolution remains
inadequate.  We, therefore, decided that only the atomic lines could
be used in modeling the observed echelle spectra.

With this limitation, the observed echelle spectra were compared to
the model spectra in two ways.  First, we examined each spectral order
by eye against the model spectra, concentrating on the depths and
wings of atomic lines, identified using a solar atlas.  We monitored,
but gave little weight to, the average strengths of the molecular
lines.  We considered only the strong atomic lines with equivalent
widths greater than 0.1 \AA, which suffer little contamination from
any blended TiO lines.  Second, we performed a least squares fit to
the strong atomic lines (giving no weight to the TiO lines).  We
extracted a small segment of spectrum centered on each atomic line
extending three line widths on both sides of line center.  In some
cases, the observed atomic lines were significantly blended with
molecular lines.  These lines were not included in the least squares
fit.  To avoid unphysical results due to degeneracy in the three free
parameters (temperature, metallicity, and surface gravity), we fixed
the metallicities of our sample stars to match the results of
\citet{Bonfils05a} listed in Table 1.  These metallicities were in all
cases consistent with our classifications by eye.

Table \ref{temperature} lists the derived temperatures from the
spectral modeling for the six M dwarfs examined here.  The first and
second columns give the star name and spectral type.
The fourth and fifth columns list our
derived values of ${\rm T_{eff}}$ from the high resolution spectra found
by eye and by least squares fitting, respectively.  For all stars but
one (GJ~424), the estimates by eye are within 60 K of the results
derived by least squares fitting, suggesting that our two fitting
procedures are self-consistent.  For GJ~424, the two methods gave
T$_{\rm eff}$ different by 150 K.  We note that the temperatures derived were
nearly the same among all echelle orders fitted for a given star, also 
suggesting self-consistency.
 
The top panel in Figure \ref{fig1} shows the best fit to GJ 436 for
the high resolution modeling (${\rm T_{eff}}$ = 3200 K, $\log$ g = 4.0,
[M/H] = 0.0), obtained both by least squares fitting and by eye.  The
observed and synthetic atomic line depths are similar in this model,
and the overall strength of the molecular lines match well.  But note
that the TiO lines that pervade the spectrum do not correspond in
detail, as discussed above.  The Lorentzian line wings of the Ca I
line at 6122 \AA \ are also well matched.  However, while the best fit
model spectrum appears to match the observed atomic lines and the
strength of the molecular lines, the surface gravity derived is
significantly lower than the known surface gravities of M dwarfs, $\log$
g $\sim$ 4.75.  This underestimate of surface gravity was consistently
observed in high resolution modeling of all the M dwarfs in our
sample, if we let $\log$ g float.  Columns 4 and 5 of Table
\ref{gravity} demonstrate this bias towards low surface gravities,
showing that in all cases the best fit to surface gravity yielded
$\log$ g $\le 4.5$.

To investigate this discrepancy in gravity further, we fixed the
surface gravity in our high resolution modeling to higher, and more
likely, values than those that appear to best fit the models.  Both
$\log$ g $ = 4.5$ and $\log$ g $ = 5.0$ were tested.  The second and
fourth panels in Figure \ref{fig1} display representative results
obtained from this test with surface gravity fixed to $\log$ g $ =
5.0$.  The strengths of the atomic lines are well-matched in the
second panel, notably those between 6135 and 6143 \AA .  However, the
Lorentzian line wings of the Ca I line at 6122 \AA \ are too pronounced.
The overall strength of the TiO lines also appear too weak in this
model, though this discrepency is more pronounced in other echelle
orders.  Similar results were observed for all stars in our sample.
Specifically, for realistic values of surface gravity, we were unable
to find models that simultaneously matched the observed atomic line
profiles and the overall strength of the TiO lines.

We believe this anomaly is due to inaccuracy in the continuous
opacity of the models.  For M dwarfs observed in the optical, the
dominant source of continuous opacity is TiO.  Problems due to
treatment of TiO are immediately seen in the lack of agreement between
the resolved lines in the observed and model spectra.  The pattern of model and
observed TiO lines do not agree, suggesting significant errors in the
model of the TiO energy levels, as noted previously by \citet{Valenti98}.  
Moreover, TiO molecules, in their
role as an effective continuous opacity source, affect the predicted
atomic line depths.  Increased continuous opacity causes the $\tau =
2/3$ surface of the star to reside higher in the star's atmosphere
where the temperature is lower.  In the LTE interpretation, the source
function is the Planck function, which is lower due to the lower
temperature.  The atomic lines form in a similarly cool region and are
thus less deep relative to the continuum.

However, increasing the surface gravity is degenerate with increasing the
continuous opacity, as larger gravity results in a larger TiO abundance.   
This effect is seen in the model spectra.  Increasing the
assumed surface gravity reduces the model atomic line depths (e.g. see 
panels 1 and 2 or 3 and 4 of Figure \ref{fig1}).  
Therefore, if the amount of continuous opacity in the models is
effectively too high, the surface gravity need not be as high in order
for the atomic line depths to match. As a result, the best fit surface
gravities will be too low.  This effect may explain the
modeling bias towards low surface gravities discussed above.

In addition, we note that if the derived surface gravities are in error, the
derived effective temperatures will also be in error \citep{Buzzoni01}.
For medium-high resolution spectral
modeling, \citet{Buzzoni01} have quantified the coupling between
errors in derived surface gravity and those in effective temperature:
\\
\[ { \Delta \log {\rm g} \over \Delta {\rm T_{eff}} } = 
1.3 \left( { 1000 \over {\rm T_{eff}} } \right) ^4  {\rm dex K^{-1}} \] \\
According to this relation, a decrease in surface gravity of 
$\sim$1.0 dex will lead 
to a decrease in effective temperature for early-type M dwarfs of 
$\sim$100 K.  Applying this result to our high resolution modeling 
results and assuming a gravity deficiency of $\Delta \log$ g 
$\sim1.0$ dex leads to an adjusted effective temperature 
of ${\rm T_{eff}} \sim 3300$ K for GJ 436.  This result is in agreement with 
the effective temperature derived with surface gravity fixed to 
$\log$ g = 5.0 dex; in that case, we found 
${\rm T_{eff}} \sim 3400$ K (see column 6 of Table \ref{temperature}).

\subsection{Low Resolution Modeling}

We also modeled a low resolution red spectrum of GJ~436 kindly
provided by D. Kirkpatrick.  This spectrum was taken on 3 December
1995 using the red channel of the double spectrograph with the 158
line mm$^{-1}$ grating on the Hale 5 m telescope with an integration
time of 5 sec.  It extends from 5140-9176 \AA \ and the resolution is
$R$ = 1140, corresponding to an instrumental profile having a FWHM of 7
\AA.  To compare the observed spectrum to the models, the spectra were
cross-correlated and matched in resolution in the same way as was done
for the high resolution spectra.  Because the individual molecular
lines are not
resolved at this resolution, the full spectrum was used in the fit.

\begin{figure}
\centerline{\scalebox{0.95}{\plotone{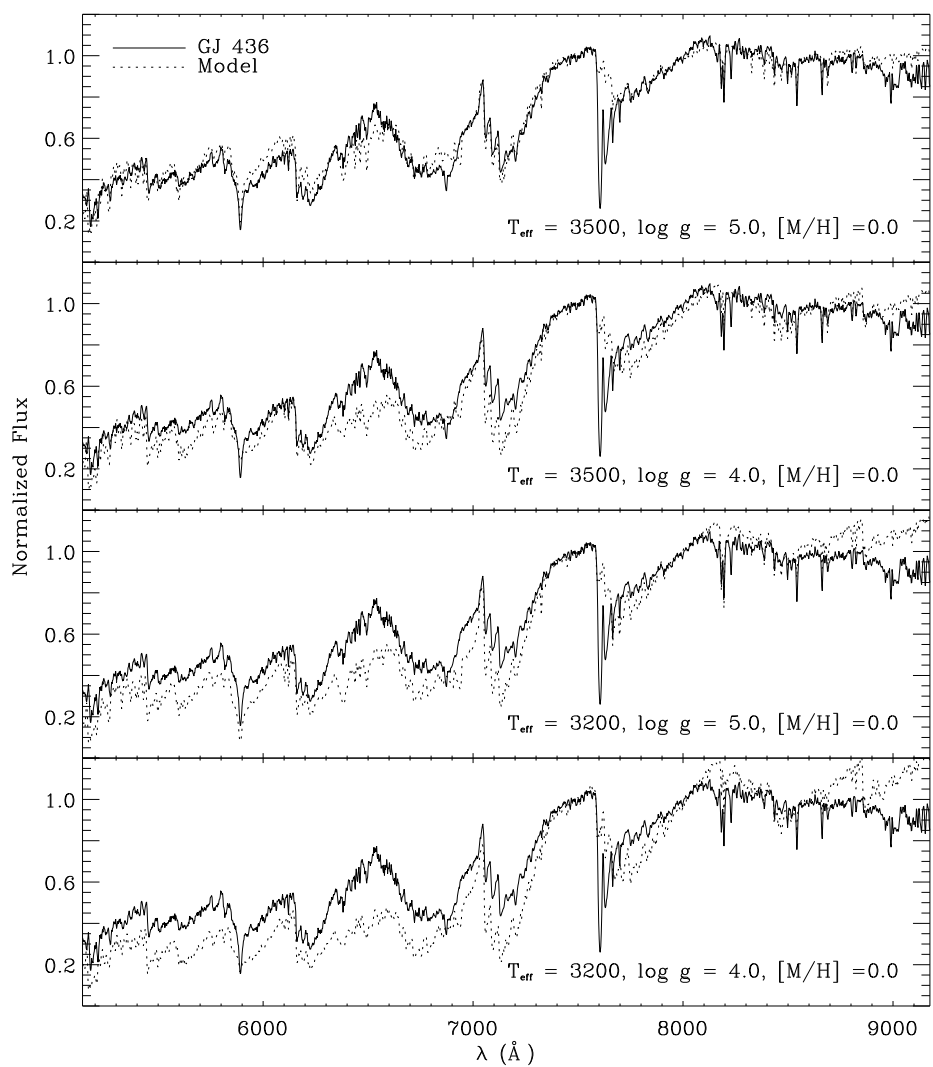}}}
\caption{Comparison, at low resolution, of the high
resolution best fit and the low resolution best fit for GJ 436.  The top
panel shows the best fit to the low resolution spectrum, while the
bottom panel shows the high resolution best fit model.  Note that at
low resolution, errors in the continuum introduced by the poorly
determined TiO lines cannot be directly observed, in contrast to the
situation at high resolution (see Figure \ref{fig1}).}
\label{fig2}
\end{figure}

The low resolution modeling of GJ 436 yielded ${\rm T_{eff}}$ = 3500 K, 
$\log$ g = 5.0, and [M/H] = 0.0.  This result is somewhat different
from the result obtained from our high resolution spectrum that 
gave ${\rm T_{eff}}$ = 3200 K, $\log$ g = 4.0, and [M/H] = 0.0.  
Figure \ref{fig2} provides a
comparison at low resolution of the best fit high resolution model to
the best fit low resolution model.  The discrepency between the results
obtained at high resolution versus low resolution is not surprising,
as the high resolution modeling is sensitive to strengths of the
atomic lines, which are influenced by the oscillator strengths
of the TiO lines.  In contrast, the low resolution modeling is
sensitive to the
shape of the continuum dictated by the gross structure of the TiO
bands.  It is interesting to note that for the three stars in our
sample that have been modeled at low resolution (GJ 436, GJ 411, and
GJ 908), the derived effective temperatures are consistently higher
than those derived at high resolution when surface gravity is left as
a free parameter. On the other hand, when surface gravity is fixed to
a reasonable value of $\log$ g = 5.0, the high and low resolution
results come into good agreement (see Table \ref{temperature}).

However, due to the inaccuracies in the TiO line list revealed by the
high resolution spectra, it is likely that the continua in the model
spectra carry significant errors at both low and high resolution.
Therefore, while the fits at low and high resolution come into good
agreement when the surface gravities are fixed to the same value, the
derived parameters may still be systematically in error.  For future
work in this area, an important distinction between modeling at high
resolution versus that at low resolution is that at high resolution,
errors in the continuous opacity due to the poorly determined TiO
lines can be directly observed in the TiO lines themselves.
At low resolution, errors in the TiO opacity are not as directly
obvious.  High resolution observational spectra are therefore required
to test new models that incorporate adjustments to the molecular
continuous opacity.  
We note that while the current TiO line list and
corresponding oscillator strengths have been adjusted and improved in
the last decade, molecular opacities remain the primary source of
uncertainty in model synthetic spectra of M dwarf atmospheres (Valenti
et al. 1998, Allard et al. 2000, Bean et al. 2006).  
The accuracy of derived parameters at the moment appears to be 
largely dependent on the observed spectral region. 
\citet{Valenti98} and \citet{Bean06}, for example, 
have improved portions of the M dwarf
models by deriving wavelengths and oscillator strengths of the TiO lines 
from the M dwarfs themselves.  As a result of this effort, \citet{Bean06}
showed that careful treatment of specific TiO bandheads
can lead to improved synthesis of the spectra of those lines. 
However, in the absence of observations containing these 
carefully tuned  regions, stellar parameters derived directly 
from synthetic spectra remain 
highly uncertain. 
The TiO opacities for the PHOENIX
models continue to be revised and tested, and improved opacities will
likely be incorporated into the next set of available grids (Allard et
al., in prep).  It would be valuable if these new grids were tested
using a sizable sample of M dwarf spectra at high resolution, as it is
likely that uncertainties in the TiO opacity have led to biases in the
currently-accepted M dwarf temperature scale.  For the present
purposes of characterizing GJ~436, however, we tentatively assign an
effective temperature that is the simple average of our low and high
resolution results: ${\rm T_{eff}} \sim 3350 \pm 300$ K.

\section{Doppler Measurements and New Orbital Model for GJ~436b}

We have obtained 59 spectra of GJ~436 at the Keck 1 telescope with the
HIRES echelle spectrometer (Vogt et al. 1994) during the 6.5--year
period, Jan 2000 to July 2006 (JD = 2451552-2453934).  These
velocities include 17 new, unpublished measurements made during the
past two seasons since announcement of the planet, GJ~436b
\citep{Butler04}.  In addition, we remeasured the Doppler shifts of
all past spectra using a newly improved Doppler analysis pipeline that  
includes a filter for the telluric absorption
lines and a superior template spectrum for spectral modeling.  The
exposure times were typically 8 min yielding S/N~$\approx$~150 and
resulting in an uncertainty in the radial velocity of 2.6 \ms (median)
per exposure.

The times of all observations, the velocities, and the uncertainties
are listed in Table 4.  Effects due to secular acceleration have been calculated 
and removed from the listed velocities.  The uncertainties consist of internal
errors only, based on the uncertainty in the mean Doppler shift of all
$\sim$700 spectral segments.  Occasionally we obtained two or three
consecutive spectra within a 30 minute interval from which we computed
the weighted average velocity and the correspondingly reduced
uncertainty.   Figure 3 shows all of the measured velocities
vs. time for GJ~436.

We attempted to fit the velocities for GJ~436 with several orbital
models shown in Figures 4-6.  A circular orbit fit to the velocities,
shown in Figure 4, yields residuals correlated in phase and a large
value of \scs = 2.04.  This fit is unacceptable, and we carry out a
statistical assessment of it in section \S 5 .  A full Keplerian
model, with the eccentricity allowed to float, produced a superior
fit, as shown in Figure 5.  This model yields residuals with RMS =
4.76 \ms and \scs = 1.69, both considerably smaller than the circular
orbit.  The best-fit parameters from this Keplerian model were $P$ =
2.6439 d, $e$=0.185, $K$=18.25 \ms .  Adopting the (revised) stellar
mass of 0.44 \msun implies a minimum mass for the planet of \msini =
0.0706 \mjup = 22.4 \mearth and a semi-major axis of 0.0285 AU.

The linear trend in the velocities evident in Figure 3 motivated a
final model that combines both a Keplerian orbit and a linear trend in
the velocities, presumably caused by a more distant orbiting
companion.  A least-squares fit to the velocities gave residuals with
RMS = 4.27 \ms, \scs = 1.57, both superior to (lower than) those from
the Keplerian model without a trend.  (The additional free parameter
for the trend was suitably included in both statistics.)  This model
with the trend gave orbital parameters, $P$=2.64385$\pm$0.00009 d,
$e$~=~0.160 $\pm$0.019, $K = 18.34 \pm 0.52$ \mse, and a linear
velocity slope of 1.36 \ms per year.  All orbital parameters are
listed in Table 5, and they are not greatly different from those
obtained with no linear trend. The new orbital parameters are only
slightly different from those in \citet{Butler04} who found $P$=2.6441
d, $K$=18.1 \ms, $e$=0.12.  But the current linear trend of 1.36 \ms
per year is smaller than that found by \citet{Butler04}: 2.7 \ms per year.  
The modest reduction in RMS and $\sqrt{\chi_{\nu}^2}$  warrants an
assessment of the reality of the trend, provided in \S 5.

We carried out all Keplerian fits by assigning weights to each Doppler
measurement that are the inverse of the quadrature sum of the internal
velocity errors and the estimated jitter, 1.9 \ms for similar M
dwarfs, based on the velocity RMS of stable M dwarfs.  The best-fit
orbital parameters are very weakly dependent on the precise value of
jitter.

The model that includes a Keplerian with a linear trend yields the
most likely physical parameters for the planet.  Adopting the
(revised) stellar mass of 0.44 \msune, the best-fit model implies a
minimum mass for the orbiting companion of \msini = 0.0713 \mjup =
22.6 \mearth and a semi-major axis of 0.0285 AU.  We note that the
value of \msini found here is higher than that (0.067 \mjup) reported
by Butler et al. (2004) due primarily to the 7\% higher stellar mass
adopted.  However, the improvements to the Doppler analysis have
reduced the RMS of the velocity residuals to the fit from 5.3 \ms to
4.3 \ms .

The non-zero eccentricity of $e$=0.16$\pm$0.02 is somewhat surprising.
Among the 23 exoplanets with periods under 4 d, this eccentricity is
the highest \citep{Butler06}, and only one other planet may have an
eccentricity as high as 0.10.  Tidal circularization is thought to be
responsible for the nearly circular orbits of the short period
planets.  If so, the high orbital eccentricity of this close-in,
Neptune-mass planet poses a mystery about its origin.  Two possible
resolutions are that a more distant planet pumps its eccentricity or
that the tidal $Q$ value is high enough to avoid tidal circularization
during the 3-10 Gyr age of this system.

\section{Orbital Constraints on GJ 436b}

Given the short orbital period of GJ 436b, a detection of a non-zero
eccentricity can carry implications for eccentricity
evolution in this system.  In this section, we evaluate the
observational evidence for the planet's non-zero eccentricity
and the presence of a long-term trend in the radial velocity data.
Because the eccentricity of a bound orbit must lie between
zero and unity, the best-fit orbit for systems with small orbital
eccentricities will suffer from a systematic Lutz-Kelker bias toward larger
eccentricities.
%
%
As noted in \S 4, the best-fit orbital solution has an eccentricity of
$e=0.160$ and bootstrap-style resampling suggests an uncertainty of
order $0.019$.  Unfortunately, the error estimates derived from
bootstrap-style resampling can significantly underestimate the true
uncertainties in orbital parameters, as demonstrated by comparisons
with Bayesian analyses (Ford 2005; Gregory 2005).
Modern computers and advanced statistical algorithms make it practical
to replace this type of frequentist analysis with a statistically
rigorous Bayesian analysis.
We perform a Bayesian analysis to determine the posterior probability
density function (posterior PDF) for the Keplerian orbital elements, 
assuming the observed radial velocity variations are due to a single
planet on a Keplerian orbit.  We assume a prior PDF that is the
product of prior PDFs for each of the model parameters individually.
We assign prior PDFs as follows: $p(P) \sim 1/P$ for $P_{\mathrm
\min}\le P \le P_{\mathrm \max}$ for the orbital period, $p(K) \sim
1/(K_o+K)$ for $K \le K_{\mathrm \max}$ for the velocity
semi-amplitude, $p(e) \sim 1$ for $0\le e <1$ for the orbital
eccentricity, $p(\omega) = 1/2\pi$ for 0$\le \omega<2\pi$ for the
argument of pericenter, $p(M_0) = 1/2\pi$ for the mean anomaly at a
specified epoch, $p(C) \sim 1$ for the mean stellar velocity, and
$p(\sigma_j) \sim 1/(\sigma_o+\sigma_j)$ for $\sigma_j \le
\sigma_{\max}$ for the stellar jitter.  In some simulations where we
also include a linear velocity trend, $D$, we have assumed a prior PDF
uniform in slope, $p(D) = 1 / 2D_{\max}$ for $-D_{\max} \le D \le
D_{\max}$.  We choose the constants $P_{\min} = 1$ d, $P_{\max} =
6.3$ yr, $K_{\max} = 2.8$ \kms, $\sigma_{\max} = 2.8$ \kms,
$K_o=\sigma_o=1$ \ms, and $D_{\max} = 10$ \ms per yr, so that the
corresponding priors can be properly normalized.  We assume that the
stellar jitter is Gaussian and uncorrelated, and we add it in quadrature
with the observational uncertainty for each observation ($\sigma_i =
\sqrt{\sigma_{\mathrm{obs},i}^2 + \sigma_j^2}$).  

The likelihood (the probability of making the actual observations for
a given set of model parameters) is computed as the product of
independent Gaussians with mean $v_{\mathrm{obs}, i}$ and standard
deviation $\sigma_i$, at each time $t_i$, using the actual observation times,
observed velocities, and uncertainties in Table 5.  We sample from the
posterior PDF using the numerical techniques of Markov chain Monte
Carlo (Ford 2005, 2006; Gregory 2005).  Fig.\ \ref{FigEbf1} (upper
panel) shows the posterior probability distribution marginalized over
all model parameters except the orbital eccentricity.  When we include
a linear slope (solid line), there is only a 5\% posterior probability
that the eccentricity is less than 0.068 and a 0.1\% posterior
probability that the eccentricity is less than 0.004, {\em if we
assume a uniform prior for eccentricity.}  Similarly, Fig.\
\ref{FigEbf1} (lower panel) shows the posterior probability
distribution marginalized over all model parameters except the slope.
When we allow for an eccentric orbit for GJ 436b (solid line), there
is a 99.8\% posterior probability that the linear slope is positive,
{\em if we assume a uniform prior for the velocity slope.}

The above analyses do not directly address the question of whether the
radial velocity observations provide evidence for a non-zero
eccentricity and/or non-zero linear slope.  To address these
questions, we must consider four sets of models: one set of models
with a planet on a circular orbit and no slope ($\mathcal{M}_{cn}$),
one set of models with a planet on an eccentric orbit and no slope
($\mathcal{M}_{en}$), one set of models with a planet on a circular
oribt and a linear slope ($\mathcal{M}_{cs}$), and one set of models
with a planet on an eccentric orbit and a linear slope
($\mathcal{M}_{es}$).  Since the $\mathcal{M}_{es}$ models have three
more model parameters than the models in $\mathcal{M}_{cn}$, we should
expect that some models from $\mathcal{M}_{es}$ will provide better
fits to the observations than the best models from $\mathcal{M}_{cn}$,
even if the planet were actually on a circular orbit.  Bayesian model
selection naturally provides a framework for quantifying the ``Occam's
razor'' factor that determines how much better the more complex model
must fit must to justify adding the extra model parameters.  We
construct a composite model ($\mathcal{M}$) that includes a discrete
indicator variable that specifies whether to use model
$\mathcal{M}_{cn}$, $\mathcal{M}_{cs}$, $\mathcal{M}_{en}$, or
$\mathcal{M}_{es}$.  We assume prior probabilities for each of
these models, $p_{cn} = p_{en} = p_{cs} = p_{es} = 0.25$.

To determine the posterior probability for an eccentric orbit, we must
compute the posterior probability PDF marginalized over all parameters
except the index specifying the model.  Unfortunately, the standard MCMC
techniques (e.g., Ford 2005, 2006) allow us to sample from the
posterior PDF assuming any one of these models, but do not provide the
normalizations.  We have used additional simulation techniques to
evaluate the ratio of the normalizations of each pair of these models.
A detailed description of the various algorithms and the advantages and
disadvantages of each will be presented in a subsequent paper (Ford et
al. 2006, in prep).  Here, we describe only one of the more
conceptually simple algorithms.  We estimate the necessary integrals
with regular Monte Carlo integration, but limit the range of integration
to the small volume of parameter space that dominates the contribution
to the marginalized posterior probability (as determined from the MCMC
simulations).  

Our Monte Carlo integration reveals the model with both
an eccentricity and a linear slope is strongly favored (by a factor of
$\simeq 10^{10}$) over the model with a circular orbit and no slope.
If we assume there is a linear slope, then the eccentric model is
$\simeq200$ times more likely than the circular model.  Similarly, if
we assume an eccentric model, then the model with a linear slope is
$\simeq 10^4$ times more likely than the model without a slope.
{\it Therefore, we conclude that the radial velocity observations provide
strong evidence for both a non-zero eccentricity and a non-zero linear
velocity trend.}

\section{Discussion}

The revised mass for the star GJ~436 of 0.44 \msun and the revised
orbital parameters and \msini for the planet have tightened the
constraints on the structure of this planetary system.  The minimum
mass of the planet remains slightly greater than that of Neptune with
\msini = 22.6 \mearth and orbiting with semimajor axis, $a$ = 0.0285
AU .

Two new results have emerged from the present analysis that render the
system interesting and puzzling.  The eccentricity is definitely
non-zero, with $e$ = 0.16$\pm$0.02, the highest eccentricity for any
exoplanet with an orbital period less than 4 d.  It has apparently
avoided tidal circularization.  Moreover, the velocities exhibit a
linear trend of 1.3 \ms per year that appears to be real, indicating
the presence of a more distant companion, the mass and orbit of which
remain poorly constrained.  It is tempting to suppose that this outer
companion is responsible for pumping the eccentricity of the inner
planet.

Two scenarios seems possible.  In the first scenario, planet b resides
in an eccentric orbit and the linear velocity slope is due to a
companion so far away that it isn't exciting the eccentricity of
planet b.  In this case, tidal theory would argue for a not-low value
of Q and hence against a rocky planet.  Scaling the Earth's Q$\sim$300
by the forcing period and scaling the Earth's radius by 22$^{1/3}$,
one gets a tidal circularization time of 2 million years, not
consistent with the age of the star that is certainly several
billion years. It shows no signs of youth such as rapid rotation, or 
enhanced magnetic, chromospheric and coronal activity.

If one scales Neptune's $Q$ value, $Q\sim 10^5$, by the forcing period
and uses its radius, then the tidal circularization time is $2
\times 10^9$ yr.  If the planet started on an eccentric orbit, then
this timescale is plausible, as the orbit will have only
partially circularized.  The $Q$ estimates for the ice giants in the
solar system range from $10^4$ to a few $10^6$.  Such values provide
an interesting constraint on $Q$ for a hot-Neptune.

In a second scenario, the planet b resides on an eccentric orbit, but the
slope is due to a planet or binary star that comes close enough to
pump the planet's eccentricity despite the tidal damping.  In this
case, planet b can be terrestrial or an ice giant since it is
continually being pumped.

But one wonders if such pumping is consistent with the velocity data.
If we naively approximate the outer ``planet c'' to be on a circular
orbit, then $m_c \sin i \sim$ 0.2 \mjup (P/20 yr)$^{4/3}$ (slope /
1.25 m/s/yr), where a 20 yr minimum orbital time scale comes from $\sim$4
times the duration of observations.
Even if we take the duration of observations to be the minimum orbital
period for planet c, then the ratio of semimajor axes is at least
$\sim$100.  With such large separation, the outer planet
is unlikely to be effective at exciting an eccentricity.

To test this, we have performed numerical integrations in the secular
octupole approximation (averaging over orbits and expanding in terms
of ratio of semi-major axes, but not eccentricities, inclinations, or
mass ratios).  We assume that this approximation gives a rough estimate of the mass
of an outer planet in a coplanar system with $\sin i$ = 1.  For an
outer planet orbital period less than 14 years (and hence masses less
than 0.12 \mjup), the outer planet would need an eccentricity larger
than 0.5 to be able to induce an eccentricity of 0.16 for planet b.
Alternatively, an outer planet with orbital period of 25 years (and
hence a mass of $\sim$0.27 \mjup) would need an eccentricity of only
0.2 to be able to induce the observed eccentricity of GJ~436b.  The
timescale for the secular eccentricity perturbations is less than
$10^6$ yr.  So this configuration would maintain the observed
eccentricity of GJ~436b, regardless of the assumed composition and
value of the tidal $Q$.  Given that we observe both an eccentricity
and a slope, this scenario offers a reasonable explanation.
Doppler observations during the upcoming years may reveal the mass and
period of the outer planet, if it exists.

\acknowledgments{The authors are grateful to D. Kirkpatrick for
supplying the low resolution spectrum of GJ 436 used in this paper.
H.M. is also grateful to D. Fischer and R. Foley for their help on
this project and to the Graduate Research Fellowship Program at NSF
for funding this research.  E.B.F. acknowledges the support of the
Miller Institute for Basic Research.  We appreciate support by NASA
grant NAG5-75005 and by NSF grant AST-0307493 (to SSV); support by NSF
grant AST-9988087, by NASA grant NAG5-12182 and travel support from
the Carnegie Institution of Washington (to RPB). We thank NASA and the
University of California for allocations of Keck telescope time toward
the planet search around M dwarfs.  This research has made use of the
Simbad database, operated at CDS, Strasbourg, France. Finally, the
authors wish to extend thanks to those of Hawaiian ancestry on whose
sacred mountain of Mauna Kea we are privileged to be guests. Without
their generous hospitality, the Keck observations presented herein
would not have been possible.}

{}

\clearpage

\begin{deluxetable}{lc}
\tabletypesize{\scriptsize}
\tablecolumns{2}
\tablewidth{0pt}
\tablecaption{Metallicities (Bonfils et al. 2005a)\label{metallicities}}
\tablehead{\colhead{Star} & \colhead{[M/H]} }
\startdata
GJ 436  & -0.03 \\
GJ 411  & -0.42 \\
GJ 424  & -0.48 \\
GJ 752A & -0.05 \\
GJ 860A & -0.02 \\
GJ 908  & -0.53 \\
\enddata
\end{deluxetable}

\begin{deluxetable}{lccccc}
\tabletypesize{\scriptsize}
\tiny
\tablecaption{\label{temperature} Summary of Effective Temperature Results}
\tablewidth{0pt}
\tablehead{
\colhead{}  &
\colhead{}  &  
\colhead{T$_{eff}$}  &
\colhead{T$_{eff}$}  &
\colhead{T$_{eff}$}  & 
\colhead{T$_{eff}$}  \\
Object & Spec. T. & Low res. & Echelle (by eye) & Echelle (num.) &
Echelle ($\log$ g = 5.0, fixed) \\
       &          & (K)      & (K)              & (K)            &
(K)}
\startdata
GJ 436  & M2.5 & 3500 & 3200 & 3200 & 3400 \\
GJ 411  & M2   & $>$3500, 3500, 4000\tablenotemark{*} 
& 3400 & 3370 & 3630 \\
GJ 424  & M0   &  -   & 3400 & 3550 & 3830 \\
GJ 752A & M2.5 &  -   & 3300 & 3240 & 3430 \\
GJ 860A & M3   &  -   & 3200 & 3140 & 3380 \\
GJ 908  & M1   & 3700\tablenotemark{+} & 3500 & 3530 & 3790 \\
\enddata
\tablenotetext{*}{Derived by Kirkpatrick (1993), Leggett et
al. (1996), and Jones et al. (1996), respectively.}
\tablenotetext{+}{Derived by Leggett et al. (1996)}
\end{deluxetable}

\begin{deluxetable}{lcccc}
\tabletypesize{\scriptsize}
\tiny
\tablecaption{\label{gravity} Summary of Surface Gravity Results}
\tablewidth{0pt}
\tablehead{
\colhead{}  &
\colhead{}  &  
\colhead{log g}  &
\colhead{log g}  &
\colhead{log g}  \\
Object & Spec. T. & Low res. & Echelle (by eye) & Echelle (num.) \\
       &          & (dex)    & (dex)	   	& (dex) }
\startdata
GJ 436  & M2.5 & 5.0 & 4.0 & 4.0 \\
GJ 411  & M2   & 5.0\tablenotemark{*} & 4.5 & 4.3 \\
GJ 424  & M0   & -   & 4.0 & 4.2 \\
GJ 752A & M2.5 & -   & 4.5 & 4.0 \\
GJ 860A & M3   & -   & 4.0 & 4.0 \\
GJ 908  & M1   & -   & 4.5 & 4.4 \\
\enddata
\tablenotetext{*}{Derived by Jones et al. (1996)}
\end{deluxetable}

\begin{deluxetable}{rrr}
\tablenum{4}
\tablecaption{Radial Velocities for GJ~436}
\label{}
\tablewidth{0pt}
\tablehead{
\colhead{JD}         & \colhead{RV}     & \colhead{Unc.}  \\
\colhead{-2440000}   & \colhead{(\ms)}  & \colhead{(\ms)} 
}
\startdata
 11552.077 &    5.84 &    2.3  \\
 11583.948 &    0.67 &    2.0  \\
 11706.865 &  -12.05 &    2.6  \\
 11983.015 &    9.48 &    2.8  \\
 12064.871 &   12.76 &    2.9  \\
 12308.084 &   19.86 &    2.5  \\
 12333.038 &  -26.89 &    3.4  \\
 12334.054 &   17.15 &    2.4  \\
 12334.935 &   -1.45 &    2.7  \\
 12363.039 &   13.43 &    2.9  \\
 12681.057 &   11.36 &    2.9  \\
 12711.898 &    0.00 &    2.4  \\
 12712.902 &    5.14 &    2.7  \\
 12804.878 &   18.86 &    2.6  \\
 12805.829 &   -7.21 &    2.4  \\
 12828.800 &   14.85 &    2.5  \\
 12832.758 &  -21.88 &    2.4  \\
 12833.763 &   13.01 &    2.4  \\
 12834.779 &   -3.56 &    3.0  \\
 12848.752 &  -21.18 &    2.6  \\
 12849.762 &   17.95 &    2.1  \\
 12850.764 &   -3.77 &    2.1  \\
 12988.146 &   -6.43 &    1.2  \\
 12989.146 &  -16.33 &    1.8  \\
 13015.141 &  -13.13 &    1.5  \\
 13016.073 &   11.10 &    1.5  \\
 13017.046 &    2.74 &    1.6  \\
 13018.142 &   -7.65 &    1.8  \\
 13044.113 &  -18.32 &    1.5  \\
 13045.018 &   -1.57 &    1.5  \\
 13045.985 &    6.91 &    1.4  \\
 13069.032 &   14.71 &    1.5  \\
 13073.991 &   -0.89 &    1.9  \\
 13077.066 &   16.30 &    2.7  \\
 13153.817 &   22.57 &    2.0  \\
 13179.759 &   -4.15 &    2.6  \\
 13180.803 &    8.51 &    2.3  \\
 13181.746 &  -11.20 &    2.0  \\
 13189.787 &  -20.17 &    1.7  \\
 13190.754 &   12.47 &    1.8  \\
 13195.767 &    0.19 &    1.7  \\
 13196.772 &   -7.03 &    2.0  \\
 13339.140 &   26.74 &    2.5  \\
 13340.129 &   -6.55 &    2.2  \\
 13370.133 &    4.71 &    2.9  \\
 13401.055 &  -12.44 &    2.5  \\
 13483.876 &    4.43 &    2.4  \\
 13693.112 &   18.49 &    2.2  \\
 13695.138 &   -3.55 &    1.8  \\
 13724.143 &   -1.82 &    2.8  \\
 13725.120 &   21.84 &    2.6  \\
 13748.059 &    3.06 &    2.4  \\
 13753.075 &   -7.75 &    2.4  \\
 13754.040 &   28.81 &    2.6  \\
 13776.052 &   -8.42 &    2.2  \\
 13777.023 &  -10.13 &    2.2  \\
 13807.020 &   21.15 &    2.5  \\
 13841.887 &    6.84 &    2.3  \\
 13933.781 &   23.54 &    2.6  \\
\enddata
\end{deluxetable}
\clearpage

\begin{deluxetable}{lr}
\tablenum{5}
\tablecaption{Orbital Parameters for GJ~436}
\label{param}
\tablewidth{0pt}
\tablehead{
\colhead{Parameter} & \colhead{} \\
}
\startdata
 $P$ (d)         		   &  2.64385 (0.00009)        \\
${\rm T}_{\rm p}$ (JD)     &  2451551.716 (0.01)  \\
$e$                          &  0.160 (0.019)         \\
$\omega$ (deg) 	           &  351 (1.2)            \\
$K_1$ (\ms)    		   &  18.34 (0.52)           \\
$f_1$(m) (M$_\odot$)       &  1.6258e-12             \\
\arel (AU)                 &  0.0285                \\
$M\sin i$ (M$_{Jup}$)      &  0.0713 (0.006)         \\
{\rm d}$v/{\rm d}t$ (\ms per yr)  &  1.36 (0.4)             \\
${\rm Nobs}$               &  59                 \\
RMS (\ms)                  &  4.27                  \\
$\sqrt{\chi^2_{\nu}}$      &  1.57                \\
\enddata
\end{deluxetable}
\clearpage

\begin{figure}
\plotone{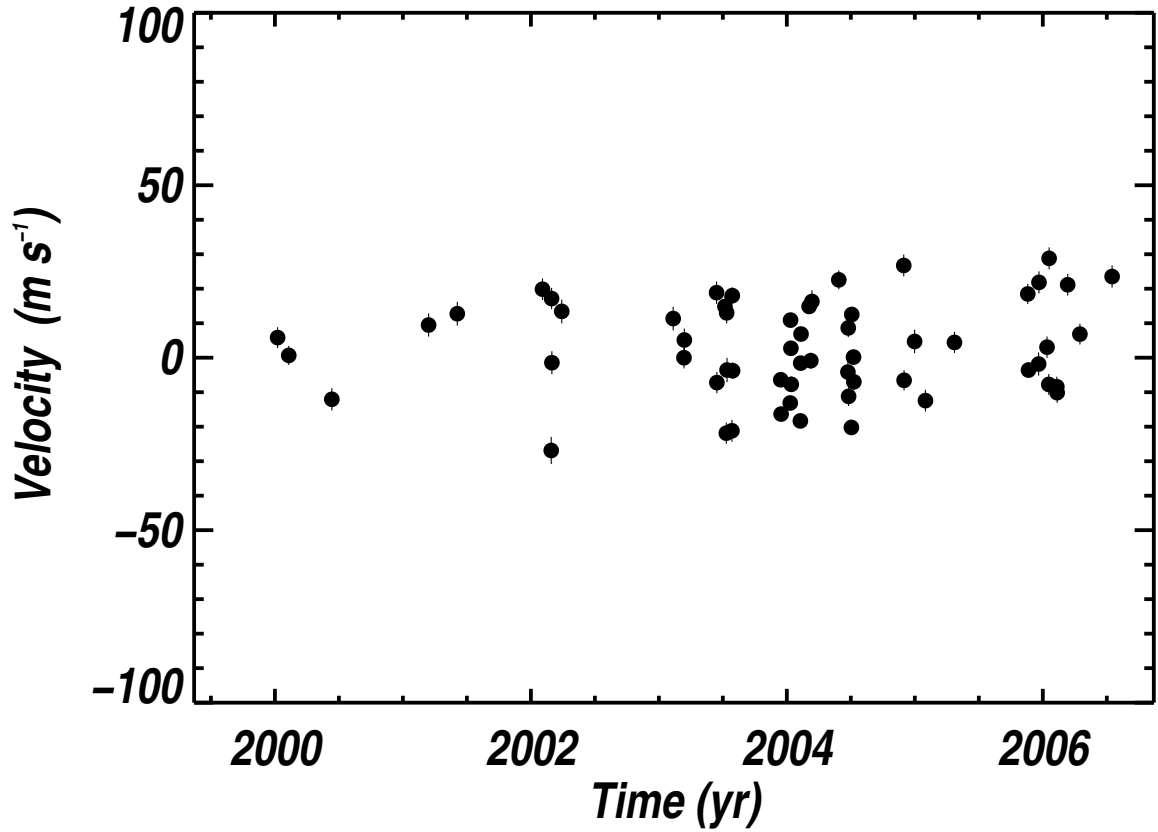}
\caption{Measured velocities vs. time for  GJ~436.   
The RMS scatter of $\sim$13.5 \ms is greater than the uncertainties
($\sim$4 \ms) shown as error bars, indicating real variation
in velocity.   There is a hint of an upward trend in the velocities.
The error bars show the quadrature sum
of the internal errors (median 2.6 \mse) and jitter (1.9 \mse). }
\label{fig3}
\end{figure}

\begin{figure}
\plotone{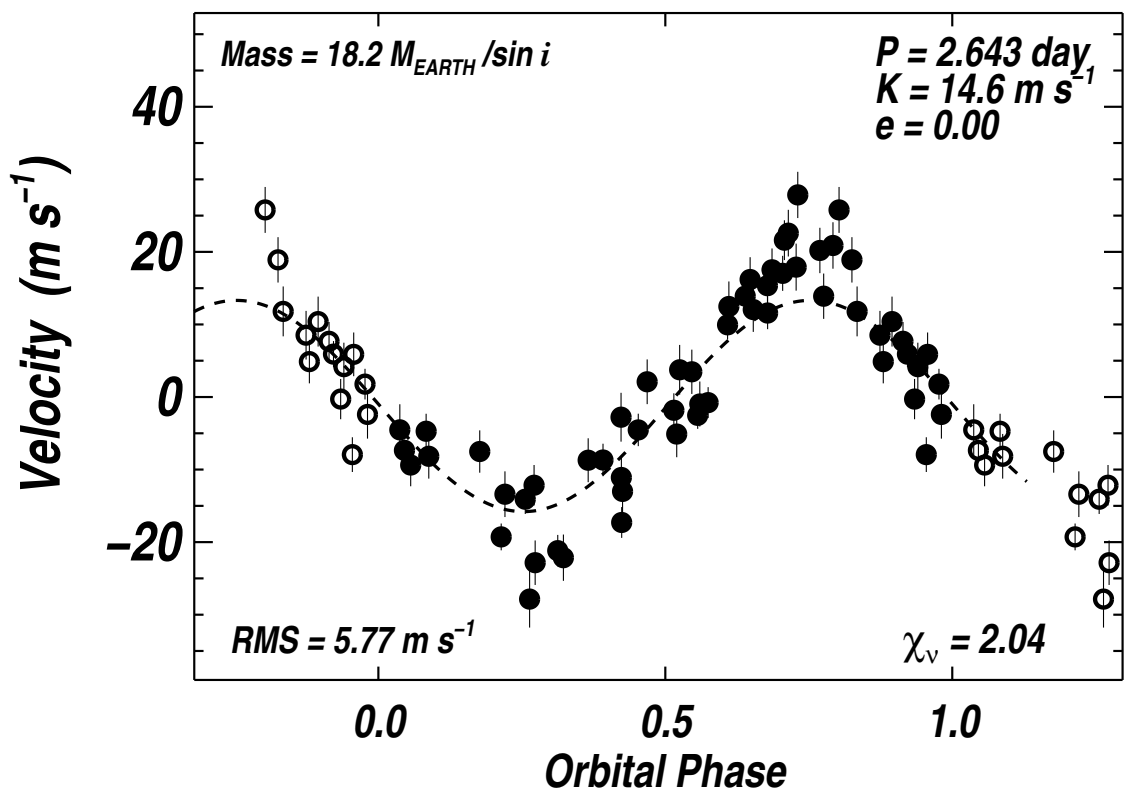}
\caption{Circular orbit fit (dashed line) to measured velocities (filled
 dots) vs. orbital phase for  GJ~436. Repeated points outside phases 0
 to 1 are shown as open circles.
The resulting parameters are:
$P$ = 2.644 d, \msini = 0.057 \mjupe = 18.1 \mearthe. 
No velocity trend was added to the Keplerian model.
The RMS of the residuals to this fit is 5.77 \ms and reduced
$\sqrt{\chi^2_{\nu}}$ = 2.04, clearly inferior
to models with non-zero eccentricity (Fig. 5, Fig. 6). }
\label{fig4}
\end{figure}

\begin{figure}
\plotone{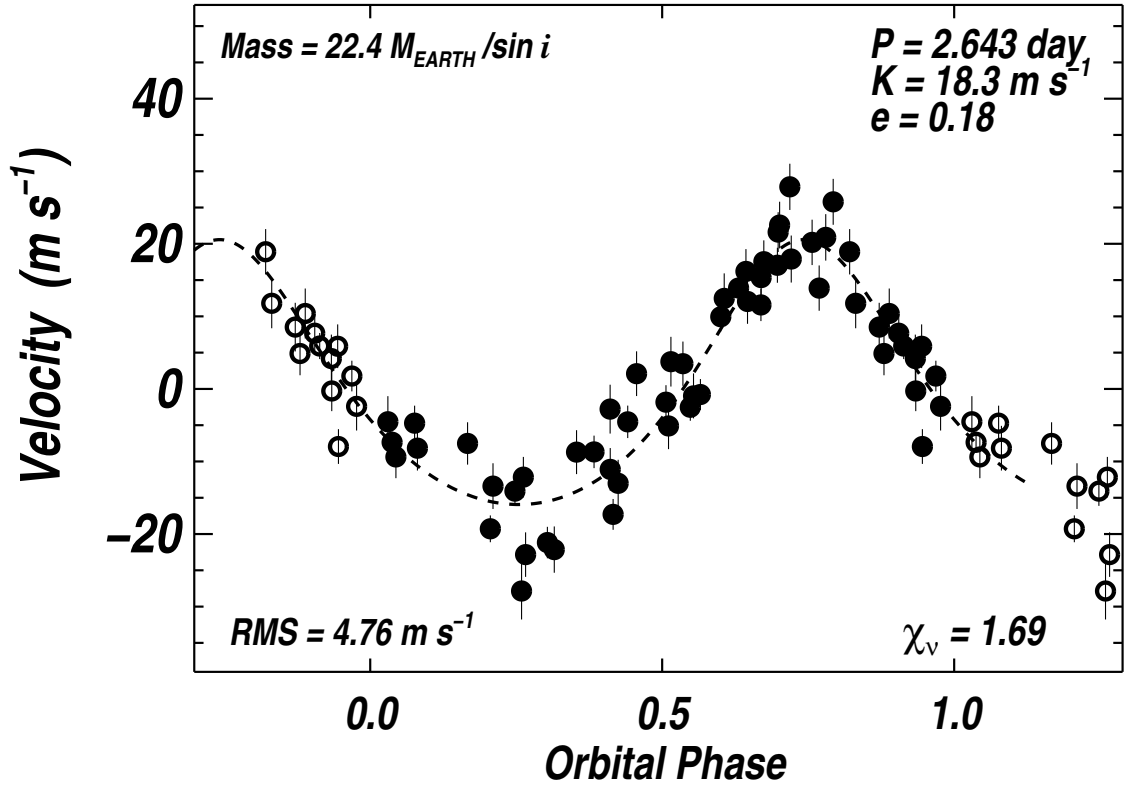}
\caption{Full Keplerian model fit (dashed line) to measured velocities
  (dots) vs. orbital phase for  GJ~436, with
repeated points (outside phases 0--1) shown as open circles.
$P$ = 2.6439 d, $e$ = 0.18, \msini = 0.0706 \mjupe =
22.4 \mearthe.  No velocity trend was added to the Keplerian model.
The RMS of the residuals to this fit is 4.76 \ms with a reduced
$\sqrt{\chi^2_{\nu}}$ = 1.69.   The uncertainties include internal errors
and jitter added in quadrature. }
\label{fig5}
\end{figure}

\begin{figure}
\plotone{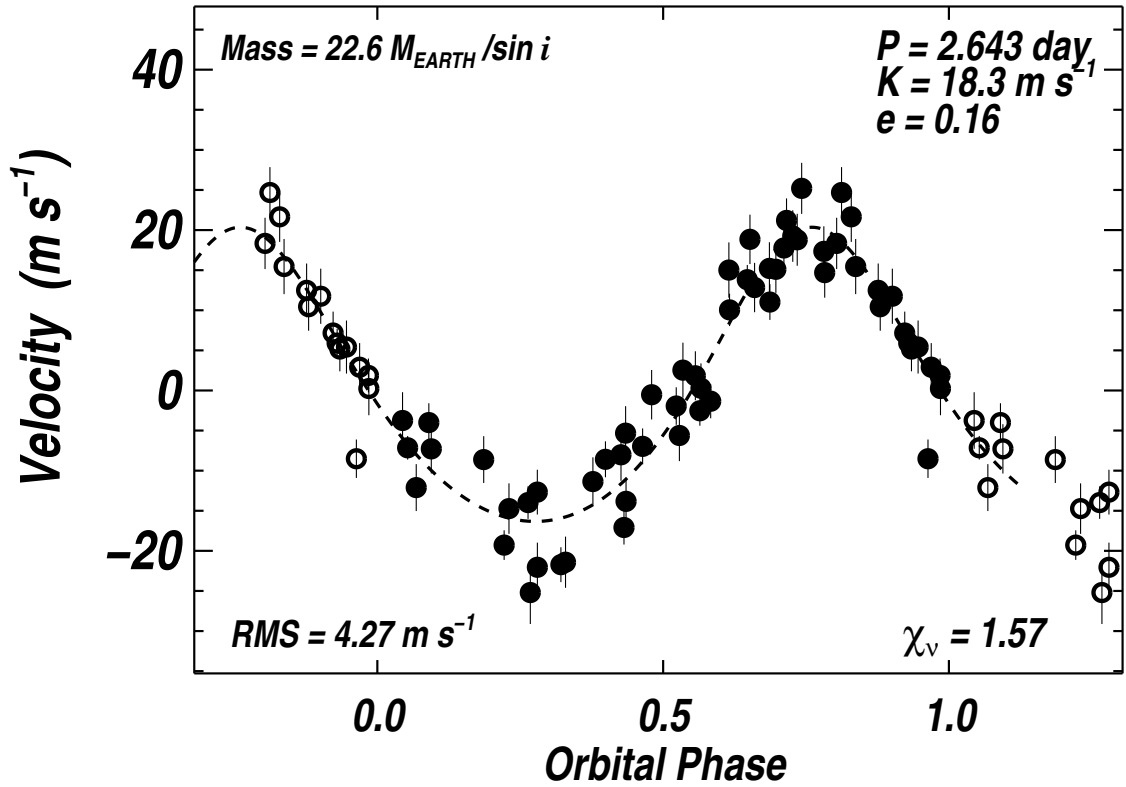}
\caption{Keplerian model plus a linear trend (dashed line) fit to measured velocities
(dots) vs. orbital phase for  GJ~436.   The best-fit orbital
  parameters are $P$ = 2.6439 d, $e$ = 0.16, \msini = 0.0713 \mjupe =
22.6 \mearthe.  This model with a linear trend gives the lowest RMS of the
residuals, 4.27 \ms, and the lowest value of
\scs=1.57.}
\label{fig6}
\end{figure}

\begin{figure}
\plotone{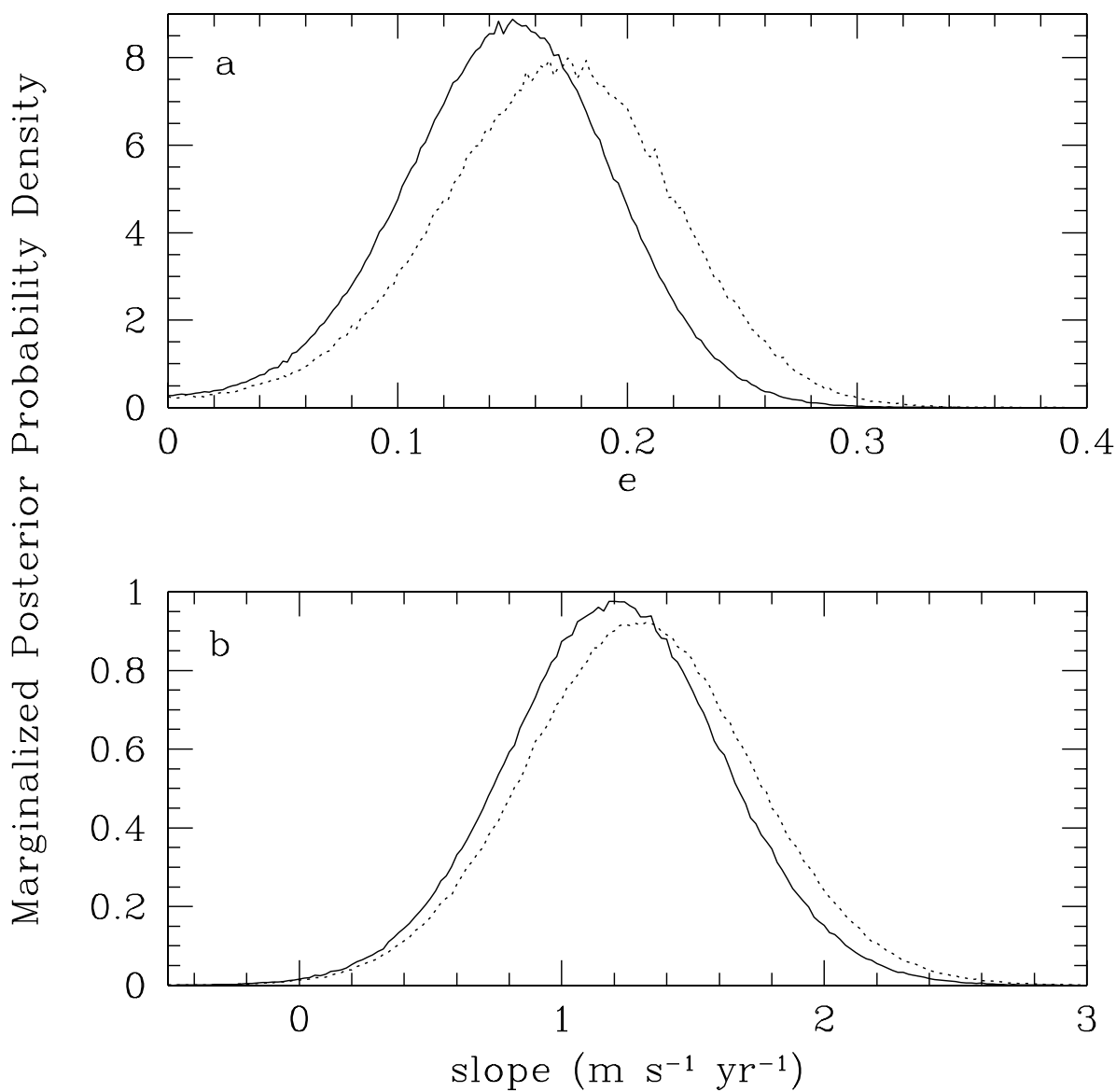}
\caption[FigEbf1]{
The upper panel shows the posterior probability distributions
marginalized over all model parameters except the orbital eccentricity
for the observations of GJ~436.  In the upper panel the solid
(dotted) curve assumes a model with (without) a linear slope.  The
lower panel shows the posterior probability distribution marginalized
over all model parameters except the slope.  In the lower panel the
solid (dotted) curve assumes a model with a slope and a single planet
on a Keplerian (circular) orbit.
\label{FigEbf1}}
\end{figure}

\end{document}